\documentclass[useAMS,usenatbib]{mn2e}
\usepackage{graphics}
\usepackage{graphicx}
\usepackage{multirow}
\usepackage{amssymb}
\usepackage{color}
\newcommand{\be}{\begin{equation}}
\newcommand{\ee}{\end{equation}}

\def \vs {{\it vs.}\ }
\def \ie {{\it i.e.}\ }

\def  \LCDM{$\Lambda$CDM}

\newcommand{\hmsun}{{\,\rm h^{-1}M}_\odot}
\newcommand{\hmpc}{{\,\rm h^{-1}Mpc}}
\def\br{{\bf r}}
\def\bg{{\bf g}}

\def\bff{{\bf f}}
\def\bv{{\bf v}}
\newcommand\apjs{ApJS}
\newcommand\apjl{ApJ}
\newcommand\aap{A\&A}
\newcommand\apj{ApJ}
\newcommand\mnras{MNRAS}
\newcommand\aj{AJ}

\begin{document}
\title{The Local Hubble Flow: Is it a Manifestation of Dark Energy? }
\author[Hoffman et al.]
{\parbox[t]\textwidth{Yehuda Hoffman$^1$, Luis A. Martinez-Vaquero$^2$, Gustavo Yepes$^2$ and Stefan  Gottl\"ober$^3$}
\vspace*{6pt} \\
$^1$Racah Institute of Physics, 
Hebrew University, 
Jerusalem 91904, Israel 
\\
$^2$Grupo de Astrof\'{\i}sica, 
Universidad Aut\'onoma de Madrid,
Madrid E-28049, Spain 
\\
$^3$Astrophysikalisches Institut Potsdam,
An der Sternwarte 16, 14482 Potsdam, Germany 
\\
}
\date{\today}
\maketitle

\begin{abstract}
  To study the local Hubble flow, we have run constrained dark matter
  (DM) simulations of the Local Group (LG) in the concordance \LCDM\
  and OCDM cosmologies, with identical cosmological parameters apart
  from the $\Lambda$ term. The simulations were performed within a
  computational box of $64\hmpc$ centred on the LG. The initial
  conditions were constrained by the observed peculiar velocities of
  galaxies and positions of X-ray nearby clusters of galaxies.  The
  simulations faithfully reproduce the nearby large scale structure,
  and in particular the Local Supercluster and the Virgo cluster.
  LG-like objects have been selected from the DM halos so as to closely
  resemble the dynamical properties of the LG.  Both the \LCDM\ and
  OCDM simulations show very similar local Hubble flow around the
  LG-like objects. It follows that, contrary to recent statements, the
  dark energy (DE) does not manifest itself in the local dynamics.
\end{abstract}


\section{Introduction}
\label{sec:intro}

It has been recently stated that the cosmological constant ($\Lambda$) or its
generalisation the dark energy (DE), manifests itself in the dynamics
of the local universe (\citeauthor{bar01} \citeyear{bar01},
\citeauthor{che04} \citeyear{che04}, \citeauthor{tee05}
\citeyear{tee05}, \citeauthor{che06} \citeyear{che06},
\citeauthor{Chernin_07b} \citeyear{Chernin_07b}, and
\citeauthor{Chernin_07a} \citeyear{Chernin_07a}). In these papers, the
coldness of the local Hubble flow around the Local Group (LG) has been
attributed to the existence DE. This has been supported by
\citet{MGH05} who analysed a set of N-body simulations and concluded
that indeed .\textit{..[their] results provide new, independent evidence for the presence of dark energy on scales of a few   megaparsecs}.  
These results,
if correct, would have provided
an independent corroboration  to  the DE component   whose existence is
otherwise inferred from observations of distant objects and the early Universe.
These authors used the term 'local' as describing the neighborhood of the LG out to a distance of a few Mpc.

 Much of the dynamical implications of the cosmological constant for the large scale
 structure were worked out by \cite{lah91}. The LG constitutes a quasi-linear object
 and therefore its dynamics cannot be modeled by the linear theory or the spherical top-hat model.
 Consequently, we have recently studied the local universe by means of constrained
 simulations (CSs, \citeauthor{kra02} \citeyear{kra02},
 \citeauthor{kly03} \citeyear{kly03}, \citeauthor{mvyh07}
 \citeyear{mvyh07} and \citeauthor{HLYD07} \citeyear{HLYD07}).  The
 unique feature of the CSs is that their initial conditions are
 generated as constrained realizations of Gaussian random fields
 \citep{HR91}. The initial conditions are constrained by observational
 data and hence they are designed to reproduce the main gross features
 of the local large scale structure. As such they provide the optimal
 tool for studying the dynamics of the LG, being a given individual but
 not an atypical object. 
 In particular the recent constrained flat $\Lambda$ dominated ( \LCDM) and
 open (OCDM) cold dark matter N-body simulations of \citet{mvyh07} were
 designed to study the local dynamics in cold dark matter cosmologies
 with and without a DE component.  These simulations are to be used
 here as a laboratory for thesting the hypothesis that the cold local
 Hubble flow is a signature of dark energy.
 We are less interested here in the actual coldness of the flow and more in the possibility 
that the DE affects the local flow. A thorough analysis of the  issue of  the coldness of the local flow is to be given elsewhere (Martinez-Vaquero et al, in preparation).
 In what follows 'local'
 is defined as the region contained in a sphere of radius $R=3$ Mpc
 centred on the LG.

 The structure of the paper is as follows. A very brief review of the
 simulations of \citet{mvyh07} is presented in Section \ref{sec:CS}.
 The selection criteria for LG candidates are summarised in Section
 \ref{sec:LG}. The flow fields around the simulated LG candidates in
 the \LCDM\ and OCDM simulations are presented in Section
 \ref{sec:Hubble}. In Section \ref{sec:grav} we compare the
 gravitational field around the LG candidates. A general discussion
 concludes the paper (Section \ref{sec:disc}).

\section{Constrained Simulations of the Local Universe}
\label{sec:CS}

 Our CSs have already been used in \citet{mvyh07} and they are briefly
 summarised here. These are dark matter (DM) only simulations employing
 a periodic cubic computational box of $64 \hmpc$ on a side using
 $256^3$ particles.  Both models use the dimensionless Hubble
 constant of $h=0.7$ (where $h=H_0/100$ km/s/Mpc), the power
 spectrum normalisation $\sigma_8=0.9$ and the cosmological matter density
 of $\Omega_m=0.3$. The \LCDM\ model corresponds to a flat universe with
 $\Omega_\Lambda=1-\Omega_m$ while for the OCDM model $\Omega_\Lambda=0$.  
These cosmological parameters correspond to the
 so-called Concordance Model. We used the parallel TREEPM N-body code
 GADGET2 (\citeauthor{gadget2}, \citeyear{gadget2}) to run these
 simulations.
For the PM part of the algorithm, we used a uniform grid  of $512^3 $
mesh points to estimate the  long-range gravitational force by means of
FFT techniques. The gravitational smoothing used to compute the
short-scale  gravitational forces  correspond to an
equivalent Plummer smoothing parameter of $\epsilon=15 h^{-1}$ kpc comoving. 
 
The number of particles used in these simulations ($256^3$) provides a
very mild mass resolution  ($1.3 \times 10^9 \hmsun$ per particle) which 
 corresponds to a minimal mass of the DM  halos of $\approx 2.5 \times 10^{10}
\hmsun$, for   objects resolved  with  more than 20 dark matter particles. 
  At such a resolution the inner
 structure of the main halos of the LG-like objects cannot be resolved,
 nor can the observed mass distribution of the LG nearby dwarfs be
 reconstructed. Yet, the dynamics on the scale of a very few Mpc is
 very well resolved.  

We set up initial conditions for these simulations in  such a way that
we can zoom in to any particular object with much more
resolution. Thus, we generate the random realizations of the density
fluctuation field  for a much larger number of particles (up to
$4096^3$). Then, we  substitute the fourier modes  corresponding to the
small wavenumber by those coming from the constrained  $256^3$ density
field and make the displacement fields according to the Zeldovich
approximation. Thus, we can now resimulate  any particular  zone 
 of the simulated  volume with   particles of variable masses,   down to
 4096 times  smaller than the particle mass  of the simulations used in
 this work.   A comparison of the results of \LCDM\  $256^3$
 simulation with   that from the  LG-like systems
 resimulated at $4096^3$ resolution does not yield 
any significant differences in their  Hubble diagrams (to be published).

 The algorithm of constrained realizations of Gaussian random fields
 (\citeauthor{HR91}, \citeyear{HR91}) has been used to set up the
 initial conditions.  The data used to constrain the initial conditions
 of the simulations is made of two kinds. The first data set is made of
 radial velocities of galaxies drawn from the MARK III \citep{mark3},
 SBF \citep{sbf01} and the \cite{kar05} catalogues. Peculiar velocities
 are less affected by non-linear effects and are used as constraints as
 if they were linear quantities (Zaroubi, Hoffman \& Dekel 1999). This
 follows the CSs performed by \cite{kra02} and \cite{kly03}. The other
 constraints are obtained from the catalog of nearby X-ray selected
 clusters of galaxies \citep{rei02}. Given the virial parameters of a
 cluster and assuming the spherical top-hat model one can derive the
 linear overdensity of the cluster. The large scale structure, \ie
 scales somewhat larger than $5\hmpc$, of the resulting density and
 velocity fields are strongly constrained by the imposed data. In
 particular all the resulting CSs are dominated by
 a Local Supercluster (LSC) - like object with a Virgo size DM halo at
 its center. The LG is not directly imposed on the initial conditions,
 but having reconstructed the actual large scale structure of the local
 universe a LG-like structure is very likely to emerge in the right
 place with dynamical properties similar to the actual ones.  The two
 simulations used here are based on the same random realization of the
 initial conditions.

\section{Selection of LG-like Candidates}
\label{sec:LG}

 The selection of LG candidates is described in detail in
 \citet{mvyh07}. The selection of the objects is based on the
 \citet{MGH05} criteria, which consist of:

 \begin{enumerate}
 \item[i.] The group contains two MW and M31 like DM halos with maximum
   circular velocity in the range of $ 125 \leq V_c \leq 270$ km/s.
 \item[ii.] The two major DM halos are separated by no more than
   $1\hmpc$.
 \item[iii.] The relative radial velocity of the two main halos is
   negative.
 \item[iv.] There are no objects with maximum circular velocity higher than MW and M31 candidates
 within a distance of  $3\hmpc$.
 \item[v.] The group resides within a distance of $5$ to $12\hmpc$ from a Virgo
   like halo of $ 500 \leq V_c \leq 1500$ km/s.
 \end{enumerate}

 DM halos are found using both the Bound Density Maxima algorithm
 (\citeauthor{bdm}, \citeyear{bdm}) and the AMIGA Halo Finder
 (\citeauthor{amiga}, \citeyear{amiga}). In the \LCDM\ simulation $26$
 LG-like objects have been found  and $43$ in the OCDM one. 
Given the fact that both
 simulation are based on the same realization of the random Gaussian
 field we have identified $9$ LG-like objects that appear in both
 simulations at about the same position and are very similar
 dynamically. We refer to these as the 'same' objects appearing in both
 simulations. 
These 'same' objects do not form any class by themselves and are statistically indistinguishable from the other LG-like objects.
 These objects are used here to exemplify the effect of
 the $\Lambda$ term on the dynamics of the LG, as they are the {\it same}
 object evolving in two identical cosmologies and environments that
 differ only by their $\Lambda$ term. 

 The fact that there is no one-to-one coincidence of the LG-like
 objects of the two simulations should not be surprising. There are two
 reasons for that. First, the two cosmologies are not identical and
 they differ in the linear gravitational growth function.  Second,
 the LG is a system in the quasi-linear regime and is far from being in
 dynamical equilibrium. Had we observed it at a slightly different time it might not be 
 qualified as  a LG-like object according to the selection criteria assumed here. 
 This is certainly the  case
for our simulated objects. Just a small miss match in the dynamical phase of the 
objects between the two simulations  can rule out an object in one or the other simulation 
from being a LG-like systems.

In the present paper we are interested in comparing the local Hubble
flow around LG-like objects. Providing that the selected systems
fulfil all requirements their exact location is not important for the
purpose of the analysis.. Therefore, we have used all the objects found
within the computational box, regardless of their position with respect
to the LSC. Some simulated LG-like objects reside close to the actual
position of the LG but they seem to be dynamically indistinguishable
from the others.

\section{The Local Hubble Flow}
\label{sec:Hubble}

\begin{figure*}

   \resizebox{13cm}{!}{\includegraphics{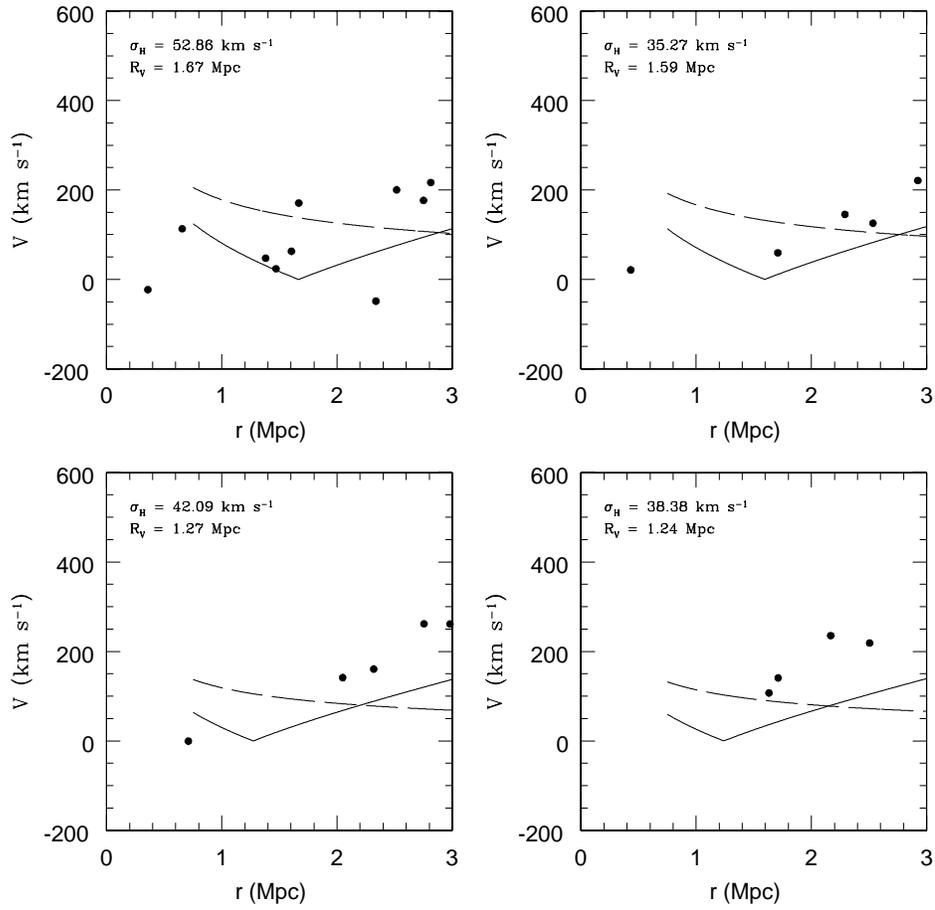}}
   \caption{ The Hubble diagrams around four LG-like objects in the
     \LCDM\ simulation. The scatter plots represent the radial peculiar
     velocity of the DM halos \vs\ the distance from the MW and
     M31-like DM halos. The solid curve corresponds to the escape
     velocity profile of the \LCDM\ model, calculated under the
     assumption of the point particle approximation. For reference the
     escape velocity of the corresponding OCDM model is given as well
     (dashed line), namely it is calculated as if the $\Lambda$ term is
     missing. The value of $\sigma_H$, 
taken over the range $0.75 \leq r \leq  3$ Mpc  
 and $R_V$ of each object is given.  }
   \label{fig:LCDM-Hubb}
 \end{figure*}

\begin{figure*}

   \resizebox{13cm}{!}{\includegraphics{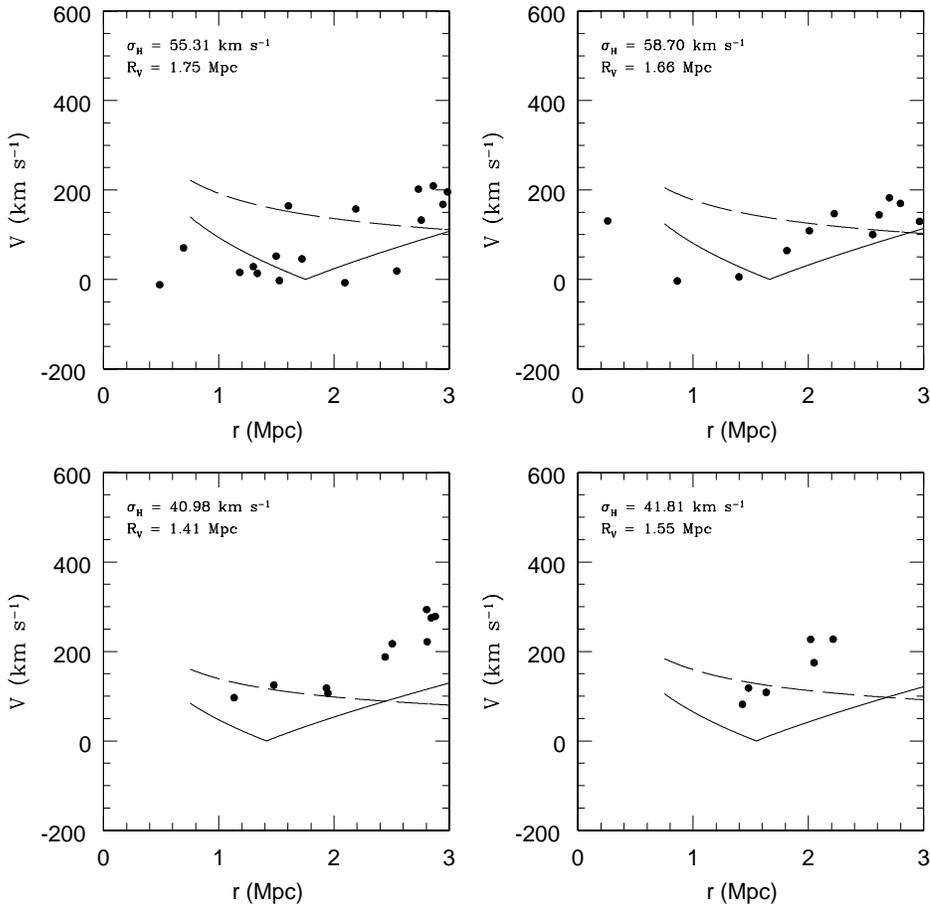}}
   \caption{ The Hubble diagrams around four LG-like objects in the
     OCDM simulation. The four LG-like objects shown here are the OCDM
     counterparts of the \LCDM\ ones shown in Fig. \ref{fig:LCDM-Hubb}.
     The structure and notations of the plots are the same as in Fig.
     \ref{fig:LCDM-Hubb}.  }
   \label{fig:OCDM-Hubb}
 \end{figure*}

The local Hubble flow around LG-like objects is probed by means of
Hubble diagrams showing the radial velocities relative to the 
objects center of mass within a distance of 3 Mpc. In Figs.
\ref{fig:LCDM-Hubb} and \ref{fig:OCDM-Hubb} we present the Hubble
diagrams of $4$ randomly chosen candidates out of the $9$ LG-like
objects which appear in both simulations (Fig. \ref{fig:LCDM-Hubb}:
\LCDM , Fig. \ref{fig:OCDM-Hubb}: OCDM). 
The figures present all the DM halos around the chosen LG-like objects out to a distance of 3 Mpc. 
A careful comparison of the
plots reveals that the Hubble diagrams of a given \LCDM\ and OCDM
simulated LG are very similar. In particular the {\it r.m.s.} value of
the scatter around a pure Hubble flow ($\sigma_H$, assuming the true value
of the Hubble constant of the simulation) does not vary statistically
between the \LCDM\ and OCDM cases. For the $4$ objects
 shown in Figs 1 and 2, 
 we find $\sigma_H=35,\
38,\ 42$ and $53$ km/s for the \LCDM\ objects and $41,\ 42,\ 55$ and
$59$ km/s in the OCDM case.
 The values of $\sigma_H$ for the other 5 common candidates are
 $31,\ 51,\ 11,\ 41$ and $79$ km/s for the objects in the \LCDM\ 
simulation and $61,\ 38,\ 18,\ 63$ and $48$ km/s in the OCDM one.

Much of the theoretical expectations for the possible manifestation of
the DE in the local flow is based on the model proposed by 
\citet[and references therein]{Chernin_07a}.
The  model essentially 
assumes that the local gravity field around the LG can be decomposed
into the contribution of the LG, modeled as a point particle, and the
contribution of the DE:
\begin{equation}
  \label{eq;g-Ch}
  g_{PP}(r)= -{G M_{LG} \over r^2} + \Omega_\Lambda H{_0^2} r
\end{equation}
The zero gravity surface is defined by $g_{PP}(R_V)=0$. The radius of
the zero gravity surface, $R_V$, plays a critical role in that simple
model. A central prediction of the model is that the local Hubble flow
should not contain galaxies with radial velocities smaller than the
escape velocity (see the Appendix), calculated under the assumption that the gravitational
field is given by the point particle approximation \citep{Chernin_07b}. This
prediction excludes galaxies residing within the LG itself, namely within
$0.7$ Mpc. To test the prediction
the radial escape velocity profiles
(Eq.  \ref{eq:vesc}) have been plotted 
 in both  Figs. \ref{fig:LCDM-Hubb}  and \ref{fig:OCDM-Hubb} as solid (\LCDM\ model) and dashed (OCDM model) lines.  From here on the term 'escape velocity' refers to the one calculated under the assumption of the point particle approximation.

 Inspection of the \LCDM\ Hubble diagram (Fig. \ref{fig:LCDM-Hubb})
 shows that indeed the prediction of \cite{Chernin_07b} is confirmed:
 only two LG-like groups  have, within the range
 $(0.7 \ - \ 3)$ Mpc,   a very few  halos each with a peculiar velocity smaller
 than the \LCDM\ escape velocity. However, this behaviour is reproduced
 by the OCDM LG-like objects equally well (Fig. \ref{fig:OCDM-Hubb}).

 To increase the statistical significance of the Hubble diagram
 analysis we have considered all the LG-like objects in the \LCDM\ and
 OCDM simulations. This is performed by plotting the radial velocities
 of all halos near the LG-like groups against their distance $r$.
 Again, the Hubble diagram of all LG-like objects in both cosmologies
 looks very similar. In fact, the fraction of halos below the escape
 velocity is somewhat  smaller in the OCDM objects  than in the \LCDM\ ones.
We conclude   that the \LCDM\ escape velocity prediction is reproduced by
 the OCDM simulation.

 The paper focuses mainly on the possible role of the DE in the dynamics of the LG. A thorough
 analysis of the coldness of the local flow will be given 
elsewhere (Martinez-Vaquero et al, in preparation).
Here a very brief summary  of the subject is given. 
The very local Hubble flow has been recently studied by \cite{kar07} and their   currently updated catalog of local peculiar velocities has been analyzed here (I. Karachentsev, private communication). Table I presents the value of $\sigma_H$ taken over 
all the DM halos (simulations) or galaxies (data) in  
the range of $[0.75 - 2]$ Mpc and $[0.75 - 3]$ Mpc of the Karachentsev's data and of the \LCDM\  and the OCDM LG-like objects. 
The cumulative distribution 
of $\sigma_H$ (calculated over the range $[0.75 - 3]$ Mpc) is presented in Fig. \ref{fig:hist}.
The plot shows that 
more than half the LG-like objects in both models  
have a $\sigma_H \leq 60$ km/s. 
So, many objects have a flow as cold, or colder, as the actual LG. Yet, as was pointed by \citet{MGH05} the real problem of the coldness lies with the relation between $\sigma_H$ and the mean density around the objects. 

It follows that the local Hubble flow around \LCDM\ and OCDM
LG-like objects is essentially indistinguishable. 
This stands 
 in clear contradiction with previous claims of \citet{bar01},  \citet{Chernin_07a} and \citet{Che07}.
Also, both the \LCDM\  and OCDM LG-like groups obey equally well the escape velocity  prediction of the flat -$\Lambda$ cosmology, as if they are not affected by  the $\Lambda$ term.

\begin{table}
   \begin{center}
     \begin{tabular}{cccc}
       \hline
       r (Mpc) & Obs.   & \LCDM\  & OCDM \\
       \hline
       $ [0.75 - 2]$   & 65  &  63     & 62 \\
       $ [0.75 - 3]$  & 68   & 72   & 75 \\
       \hline
     \end{tabular} 
     \caption{
      The value of $\sigma_H$ (in units of km/s) of the LG, compiled from the 
      Karachentsev data, and of the \LCDM\ and OCDM 
       LG-like objects combined together, in the manner of Fig.
       \ref{fig:All-Hubb}. Two distance cuts are used for calculating $\sigma_H$. }
   \end{center}
   \label{table:sigma-H}
 \end{table}

\section{The Local Gravitational Field}
\label{sec:grav}

To understand the possible reason for the discrepancies between the
present results and the 
and the model predictions 
we have studied the nature of the local gravitational field.
The prime motivation here is to check the validity of the \citet{Chernin_07a} model of the gravitational field (Eq. \ref{eq;g-Ch}), which    corresponds to the full gravitational field expressed in physical, and not co-moving, coordinates. The relation between the peculiar gravity (output of GADGET) and the physical one was derived by 
\citet{mvyh07}, but it is repeated here for the sake of completeness.

The physical $\textbf{r}$ and comoving $\textbf{x}$ coordinates are
related by: $\textbf{r} = a\textbf{x}$. The gravitational field equals
the physical acceleration of an object $\ddot{\textbf{r} }= \bg$.  The
GADGET code provides an acceleration-like term defined as:
\begin{equation}
  \bff_p=\frac{1}{a} \frac{d}{dt}\left( a\cdot \bv_p\right)=\dot{\bv}_p + H\cdot \bv_p,
  \label{fpdef}
\end{equation}
where $\bv_p=\dot\br - H\br$ is the peculiar velocity, $H$ is Hubble's
constant and $a$ is the expansion scale factor.  It follows that
\begin{equation}
  \ddot \br = \bff_p + \br \frac{\ddot a}{a} = \bff_p + \left( -\frac{1}{2}\Omega_M + \Omega_\Lambda \right) H^2 \textbf{r}.
  \label{r_f_a}
\end{equation}
Namely, the linear term corresponds to the unperturbed Friedman
solution and $\bff_p$ to the fluctuating component\footnote{Note the
  typo in Eq. A3 of \citet{mvyh07} where $\Omega_\Lambda$ was omitted. However,
  this did not affect the result, as only the fluctuating term of $g$
  was considered there.}.

The gravitational field is taken with respect to the LG
reference frame.  So that, one finally obtain:
\begin{equation}
  g= (\bff_p - \bff_p^{LG} )\frac{{\textbf{r}}}{r}+ \left( -\frac{1}{2}\Omega_M + \Omega_\Lambda \right) H^2 r
  \label{eq:g}
\end{equation}
where $r$ is the distance from the 
center of mass of the 
LG.  This is the
field acting on each dark matter particle. The total
acceleration of halos was computed by averaging this quantity over
all
particles belonging to each halo. 
The acceleration is scaled by  $H_0^2 \times 1 \,\rm $ Mpc. In such scaling
the unperturbed gravitational acceleration of a shell of radius $1
\,\rm $ Mpc equals to $-q_0$, where $q_0$ is the 
cosmological  deceleration parameter.

In Figs. \ref {fig:LCDM-g} and \ref {fig:OCDM-g} we compare  the radial
component of the exact (\ie in the sense of the simulations)
gravitational field
with the \citet{Chernin_07a} model ($g_{PP}$),
as traced by the DM halos around the LG-like objects. 
Only the fluctuating component of the gravitational field is shown in the figures, namely 
for the numerically exact field it is the radial component of $\bff_p$ and for the point particle approximation it is 
$f_{p,PP} = -G M_{LG} /  r^2  + \Omega_M H^2 r /2$. 
 We show all DM halos in the distance range of $0.7 \leq r \leq 3.0 $ Mpc 
from the same LG-like objects
that were 
presented in  Figs. \ref{fig:LCDM-Hubb} and \ref{fig:OCDM-Hubb}. 
Since DM halos closer that $r=0.7 $ Mpc are affected by the two-body dynamics
of the LG, they are excluded here.  
Fig. \ref{fig:g-g} shows the   scatter plot of the gravitation field of all the LG-like objects of the \LCDM\ and OCDM simulations.  The plots show that the numerically exact value of the radial component of $\bff_p$ and the point particle  prediction ($f_{p,PP}$) are very poorly correlated. 
A linear regression analysis finds a correlation coefficient of $0.40$ ($0.23$) and a slope of $0.35$ ($0.16$) for the \LCDM\ (OCDM) model. 
It is clear that the point particle model fails to reproduce the actual gravity field,
and therefore it cannot account for the dynamics of the LG.

\section{Discussion}
\label{sec:disc}

The most striking result of this paper is that the local Hubble flow
around LG-like objects in the OCDM model is almost indistinguishable
from the \LCDM\ flow. To the extent that the models do differ it is the
OCDM model that has somewhat colder Hubble flow than the \LCDM\ one. It
follows that the local flow is not affected by the DE and does not
manifest the present epoch dominance of the DE.

One should not be surprised by the departure of the simulated local
Hubble flow from the prediction of the simple model proposed by
\cite{Chernin_07a}. First, the actual gravitational field deviates
considerably from the predicted one.
Second, the gravitational dynamics is not local and the tidal field
plays an important role in the quasi-linear regime (\citeauthor{YH86}
\citeyear{YH86}, \citeauthor{YH89} \citeyear{YH89}, \citeauthor{zar93}
\citeyear{zar93}, \citeauthor{rien94} \citeyear{rien94},
\citeauthor{delpopo1} \citeyear{delpopo1}). It follows that the
dynamics depends not only on  the local field, but is also  affected by the
shear, namely the tidal field. The shear breaks the simple linear
relation of the density and velocity fields of the linear regime and
therefore the local density field cannot account for the local Hubble
flow. 

The comparison of the \LCDM\ and OCDM simulations shows that they yield
very similar LG-like objects with virtually identical local Hubble
flows. It follows that the dynamical properties of LG-like objects and
their environments, in the linear and quasi-linear regime, depend
mostly on the cold matter content of the universe, namely $\Omega_m$, and
only weakly on the DE. This is another manifestation of the fact that
the properties of the cosmic web, expressed in co-moving coordinates,
depend mostly on $\Omega_m$ and hardly on the DE \citep{HLYD07}.

\section*{Acknowledgements}
Fruitful discussions with A. Chernin, 
A. Macci{\`o} and A. Tikhonov are gratefully acknowledged.
We thank I. Karachentsev for providing us his updated catalog of peculiar velocities 
of galaxies in  the Local Volume.
We thank  DEISA consortium for granting us computing time in  the
MareNostrum supercomputer at  BSC (Spain)  and the SGI-ALTIX supercomputer
at LRZ (Germany)  through the Extreme Computing Project (DECI) SIMU-LU.
We also thank   NIC J\"ulich (Germany) for the access to the IBM-Regatta p690+ JUMP
supercomputer and CesViMa (Spain) for access to the Magerit IBM-BladeServer
supercomputer.   GY would like to thank also MEC (Spain) for financial support
under project numbers FPA2006-01105 and AYA2006-15492-C03.  LAMV
acknowledges financial support from Comunidad de Madrid through a PhD
fellowship.
The support of the ISF-143/02 and the Sheinborn Foundation (YH) and the
European Science Foundation through the
ASTROSIM Exchange Visits  Programme (SG) is also acknowledged.

\section*{Appendix}
Assuming the local gravitational field around the LG is indeed given by
the point particle approximation (Eq. \ref{eq;g-Ch}), the effective
Newtonian potential is given by
\begin{equation}
  \label{eq:phi}
  \phi(r) =  - { G M_{LG} \over r} - {\Omega_\Lambda H{_0^2} r^2 \over 2}.
\end{equation}
The effective Newtonian energy (per unit mass) is simply given as $\epsilon
=v^2/2 +\phi(r)$.  In the presence of the $\Lambda$ term the potential reaches a
maximum at $R_V$ at which the potential peaks at $\epsilon_V \equiv \phi(R_V)$. It
follows that a particle is unbound to the LG if its energy is larger
than $\epsilon_V$.  The escape velocity is therefore given by
\begin{equation}
  \label{eq:vesc}
  {v{^2_{esc}}\over 2} = G M_{LG}\big({1\over r} -{1\over R_V}\big) 
  +{\Omega_\Lambda H{_0^2} \over 2}\big( r^2 - R{_V^2}  \big).
\end{equation}
In the case of a vanishing cosmological constant the classical
expression of Eq. \ref{eq:phi} and \ref{eq:vesc} is recovered upon
substituting $R_V=\infty$
and $\Omega_\Lambda=0$.
 It should be reemphasised here that the present derivation is done under the assumption of the point particle approximation. The analysis of the simulation proves that the assumption is inapplicable to the LG systems.

 \begin{figure*}

   \resizebox{13cm}{!}{\includegraphics{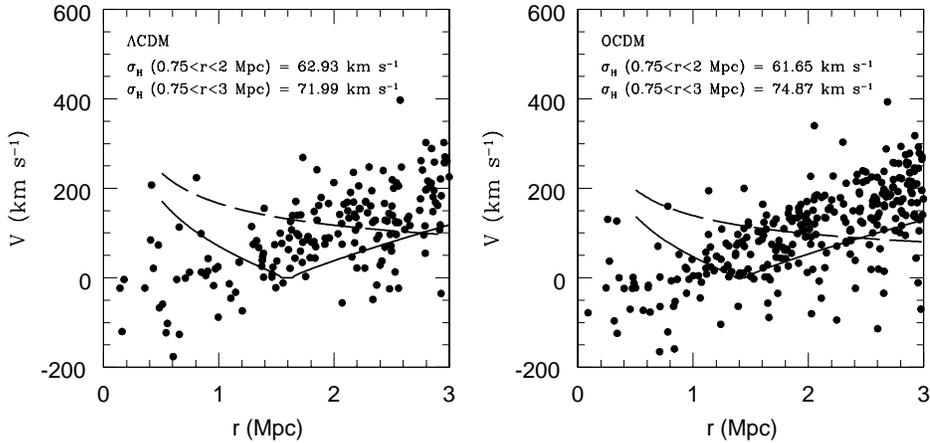}}
   \caption{ Combined Hubble diagram of 26 LG candidates in the \LCDM\
     (left panel) and 43 candidates in the OCDM (right panel) models.
     The radial distance $r$ is scaled by the value of $R_V$ of each
     object. The values of $\sigma_H$ within $2$ and $3 $ Mpc are given in
     Table 1. The escape velocity curves are plotted in the same manner
     as in Figs. \ref {fig:LCDM-Hubb} and \ref{fig:OCDM-Hubb}. $<R_V>$
     is the mean $R_V$ of all the LG-like objects for each simulation.
   }
   \label{fig:All-Hubb}
 \end{figure*}

\begin{figure*}
\resizebox{8cm}{!}{\includegraphics{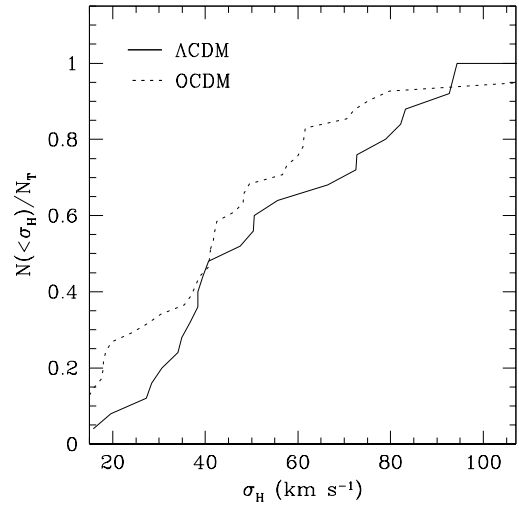}}
\caption{ 
The fractional cumulative  $\sigma_H$ function of LG-like objects,  namely the fraction of objects with $\sigma_H$ lower than a certain value. Full line corresponds to the 26 \LCDM\ objects and the dashed one to the 43 OCDM objects. The dispersion around a pure Hubble flow, $\sigma_H$,  is calculated over the range $0.75 \leq  r \leq 3 $ Mpc.
}
\label{fig:hist}
\end{figure*}

 \begin{figure*}

   \resizebox{13cm}{!}{\includegraphics{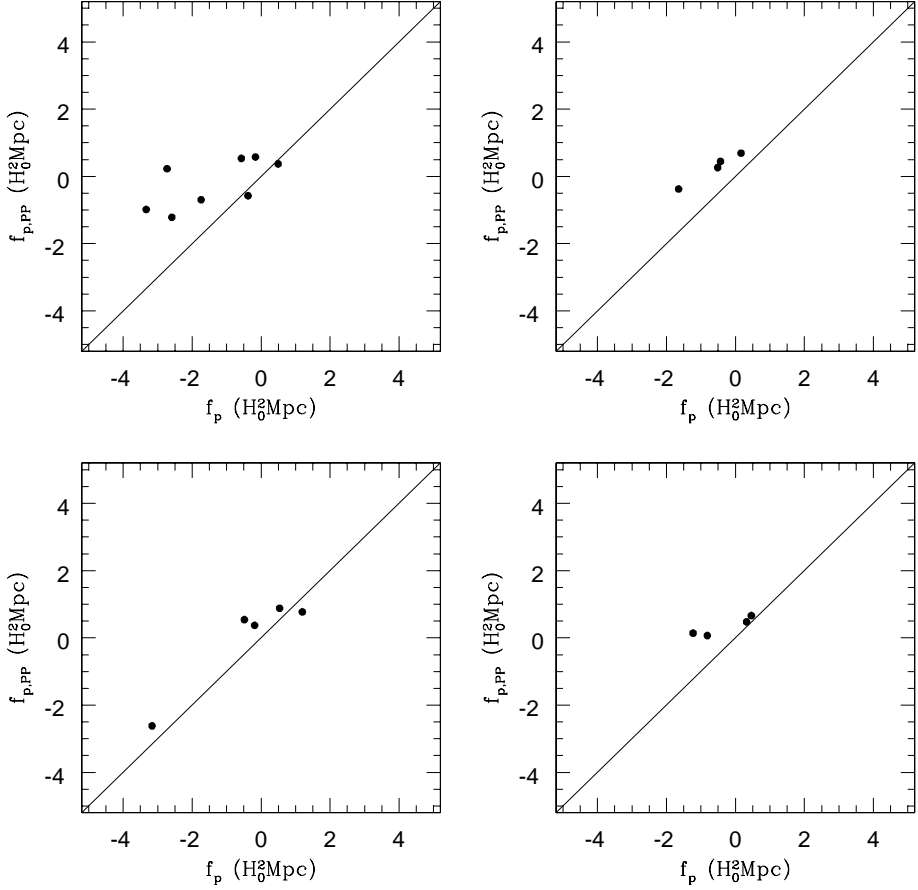}}
   \caption{ The \LCDM\ gravitational field: A scatter plot of the
     exact gravitational field $g$ (Eq. \ref{eq:g}) \vs\ the
     approximated one $g_{PP}$ (Eq. \ref{eq;g-Ch}) experienced by DM
     halos around LG-like objects. The four objects and panels
     correspond to the ones in Fig. \ref{fig:LCDM-Hubb}. Only DM halos
     in the range of $0.75 \leq r \leq 3.0 $ Mpc are plotted here. The
     gravitational field is scaled by $H_0^2 \times 1 \,\rm $ Mpc.  }
   \label{fig:LCDM-g}
 \end{figure*}

 \begin{figure*}

   \resizebox{13cm}{!}{\includegraphics{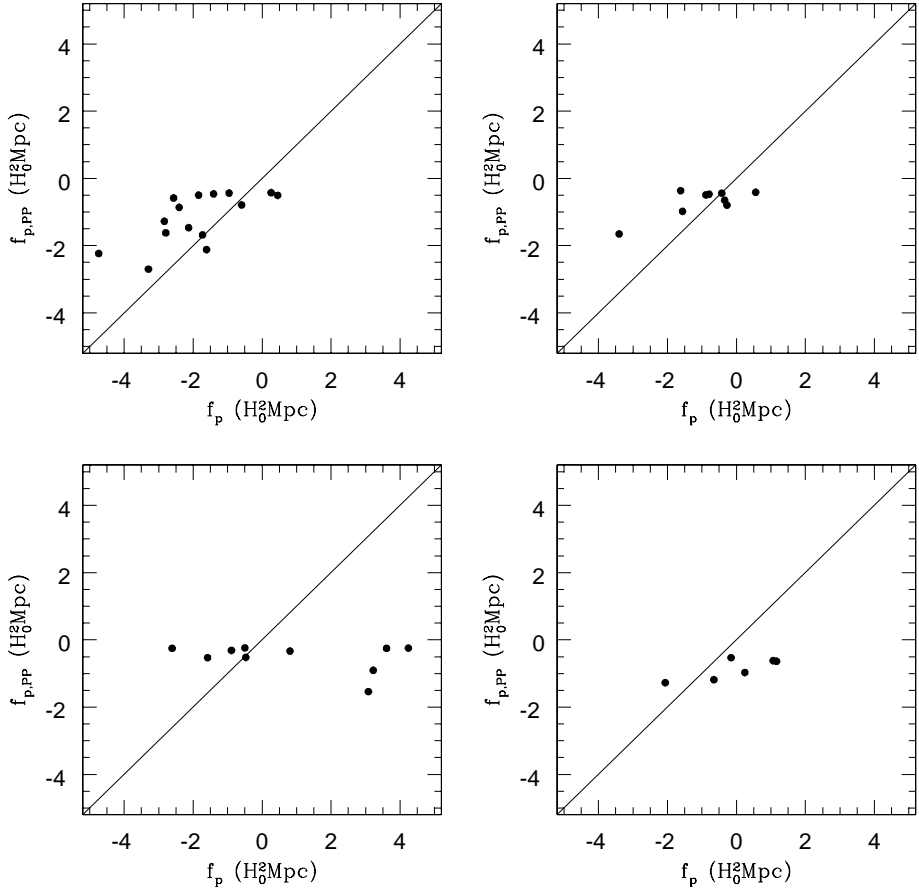}}
   \caption{ Same as Fig. \ref{fig:LCDM-g} but for the OCDM simulation.
     The LG-like objects and panels correspond to the ones in Fig.
     \ref{fig:OCDM-Hubb}.  }
   \label{fig:OCDM-g}
 \end{figure*}

\begin{figure*}

\resizebox{13cm}{!}{\includegraphics{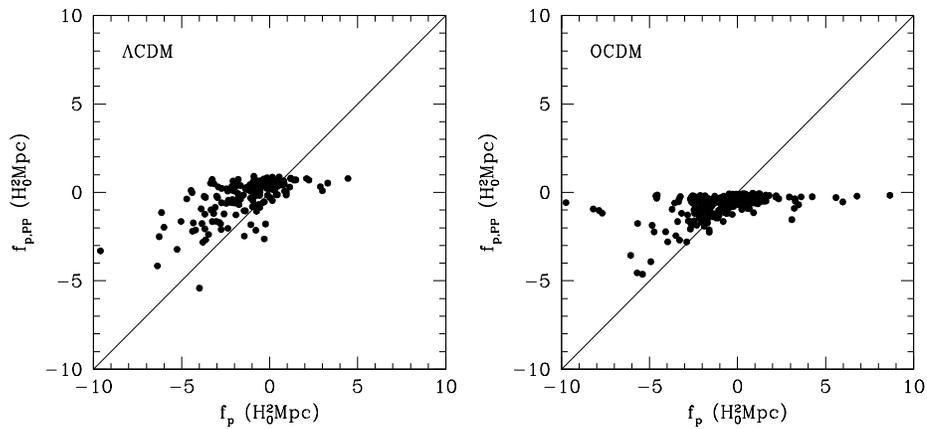}}
 \caption{
Scatter plot of the  fluctuating component of the  gravitational field predicted by the point particle model ($f_{p,PP}$) against the exact  value calculated by the simulations. The left panel  shows the distribution of the DM halos in the vicinity  ($0.75 \leq r \leq 3.0 $ Mpc) of the 26 LG-like \LCDM\ objects and the right panel exhibits the 43 OCDM objects. 
The gravitational field is scaled by $H_0^2 \times 1 $ Mpc.
}
   \label{fig:g-g}
 \end{figure*}

\end{document}